%% file: Delay_SCN_Final.tex
\documentclass[journal,twocolumn,10pt]{IEEEtran}
\usepackage{graphicx}
\usepackage{amssymb}
\usepackage{amsmath}
\usepackage{mathtools}
\usepackage{dsfont}
\usepackage{cite}
\usepackage{stfloats}
\usepackage{subfigure}
\usepackage{psfrag}
\usepackage[mathscr]{euscript}
\usepackage{acronym}  
\usepackage{booktabs}
\usepackage{bbm}

\usepackage{float}
\usepackage{balance}
\input{SupportDocuments/aconym}
\input{SupportDocuments/defmetric}

\newcommand{\paperTitle}{Delay Analysis of Random Scheduling and Round Robin in Small Cell Networks}

\begin{document}

{
\title{\paperTitle}

\author{

	    Howard~H.~Yang,
        Ying Wang,
        and Tony~Q.~S.~Quek

\thanks{Manuscript received Feb. 03, 2018, revised Mar. 31, and May 28, 2018, and accepted May 29, 2018. The associate editor coordinating the review of this letter and
    approving it for publication was Dr. Chun Tung Chou.

    This work was supported in part by the MOE ARF Tier 2 under Grant MOE2015-T2-2-104 and in part by the SUTD-ZJU Research Collaboration under Grant SUTD-ZJU/RES/01/2016.}

\thanks{H.~H.~Yang and T.~Q.~S.~Quek are with the Singapore University of Technology and Design (e-mail: howard\_yang@sutd.edu.sg, tonyquek@sutd.edu.sg). Y.~Wang is with Nanjing University of Post and Telecommunications (e-mail: wangy1585@163.com). }
}
\maketitle
\acresetall
\thispagestyle{empty}
\begin{abstract}
We analyze the delay performance of small cell networks operating under random scheduling (RS) and round robin (RR) protocols.
Based on stochastic geometry and queuing theory, we derive accurate and tractable expressions for the distribution of mean delay, which accounts for the impact of random traffic arrivals, queuing interactions, and failed packet retransmissions.
Our analysis asserts that RR outperforms RS in terms of mean delay, regardless of traffic statistic. Moreover, the gain from RR is more pronounced in the presence of heavy traffic, which confirms the importance of accounting fairness in the design of scheduling policy.
We also find that constrained on the same delay outage probability, RR is able to support more user equipments (UEs) than that of RS, demonstrating it as an appropriate candidate for the traffic scheduling policy of internet-of-things (IoT) network.
\end{abstract}

\acresetall

\section{Introduction}\label{sec:intro}
The paradigm shift from traditional macro cellular network to small cell networks is unrelenting, whereas more and more user equipments (UEs), including smartphones, tablets, and smartwatches, will be connected to the small access points (SAPs) \cite{AndBuzCho:14}.
Since the wireless spectrum is usually limited, there is inevitable contention among UEs for communication resource. At this stage, operators need to choose an appropriate traffic scheduling protocol to allocate resource among different UEs so as to meet the delay requirement \cite{AndBuzCho:14}.
Among the large number of proposed scheduling schemes \cite{gerla1980flow}, random scheduling (RS) and round robin (RR) are the most popular practical choices due to their low-implementation cost \cite{Hah:91}.
Even though, a full understanding on the respective delay performance of RS and RR is still essential to help devise insights and perform suitable design.
While the delay performance of both schemes has been well understood in wired data networks \cite{Hah:91}, it remains in darkness from the perspective of wireless small cell networks because: $i$) the irregular deployment of SAPs leads to asymmetric UE performance, e.g., some UEs may succeed in capturing a large portion of resources than others because of their relative position in the network, and thus enjoy preferential treatment;
$ii$) the shared nature of wireless channel results in the queuing evolution of any typical cell strongly interacted with its neighbors, which piles on analytical complications.

Most of the previous works only focus on one aspect of the issue \cite{Hah:91,ZhaQueHua:16,DinLopJaf:17,JafLopDin:15}, i.e., they model only the temporal arrival of packets \cite{Hah:91} or the spatial lotations \cite{ZhaQueHua:16,DinLopJaf:17,JafLopDin:15}, and thus fail to track the other. Recently, several attempts have been made to analyze delay in large-scale networks by modeling the spatial-temporal randomness \cite{ZhoQueGe:16,YanQue:18,GhaElsBad:17,GhaElSBad:17ICC}.
However, the results in \cite{ZhoQueGe:16} only give upper and lower bounds, that can be loose in light traffic condition, for the delay distribution, hence may fail to explicitly capture the interplay between spatial topology and temporal queuing evolution. While \cite{YanQue:18} attains a precise expression for the distribution of mean delay, the result holds for single UE per cell scenario, which do not allow one to investigate the effect of different scheduling policies.
The model in \cite{GhaElsBad:17,GhaElSBad:17ICC} consider uplink random access for multi-UE per cell and accounts for the intra-cell and inter-cell interference due to the scheduling policy. However, the results only analyze the RS and do not extend to other scheduling methods.

In this work, we take a complete treatment to the delay analysis in small cell networks under different traffic scheduling policies. Specifically, we model the locations of SAPs as homogeneous Poisson point process (PPP) where each cell has multiple associated UEs. The traffic arrivals are modeled as independent Bernoulli processes, and account for the retransmission of unsuccessfully delivered packets. By combining stochastic geometry with queueing theory, we derive accurate expressions for the mean delay distribution under both RS and RR protocols, allowing to compare them and to draw insightful conclusions.

\section{System Model}\label{sec:sysmod}
\begin{figure}[t!]
  \centering{}

    {\includegraphics[width=0.90\columnwidth]{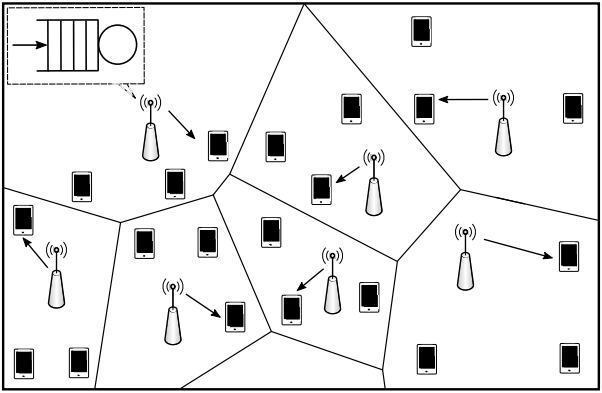}}

  \caption{ Example of a small cell network with random traffic arrivals. }
  \label{fig:SysMod_M1}
\end{figure}

We consider the downlink of a small cell network, as depicted in Fig.~\ref{fig:SysMod_M1}, that consists of randomly deployed {SAP}s.
We model the spatial locations of SAPs as a homogeneous {PPP} $\Phi_{\mathrm{s}}$  with spatial density $\LS$.
For notational simplicity, we assume each SAP has $K_{\mathrm{s}}$ associated UEs, who are uniformly distributed within the Voronoi cell it generated.\footnote{Note that the results obtained through the machinery of queueing theory and stochastic geometry can be modified to model random number of UEs per SAP \cite{yang2017packet}.}
All  SAPs and UEs are equipped with single antenna, and every SAP transmits with power $\PST$.
We assume the channels between any pair of nodes to be narrowband and affected by two attenuation components, namely small-scale Rayleigh fading, and large-scale path loss.

We use a discrete time queuing system to model the traffic profile, and segment the time axis into slots with equal duration. We assume all queuing activities, i.e., packet arrivals and departures, take place at each time slot. For a generic UE, we model its packet arrivals as independent Bernoulli processes with rates $\xi \in [0,1]$ (packet/slot) \cite{ZhoQueGe:16,GhaElsBad:17,YanQue:18}.
We further assume that each node accumulates all incoming packets in an infinite-size buffer.
At the beginning of any typical time slot, every node with a non-empty buffer initiates a transmission attempt. If the received signal-to-interference ratio (SIR) exceeds a predefined threshold, the transmission is successful and the packet can be removed from the queue, otherwise, the transmission fails and the packet remains in the buffer. We assume the feedback of each transmission, either success or fail, can be instantaneously aware by the SAPs such that they are able to schedule transmission at next time slot. Moreover, for those SAPs with empty buffers, they mute the transmissions to reduce power consumption and inter-cell interference.

In the following, we investigate two traffic scheduling protocols, i.e., random scheduling (RS) and round robin (RR), described as below \cite{ZhoQueGe:16}\footnote{For tractability purpose, the scheduling policy is applied with respect to all the UEs regardless of their queueing status \cite{ZhoQueGe:16}. In this fashion, the least amount of resource will be allocated to the active UEs (i.e., those UEs with non-empty queues) and the results derived in this paper are essentially upper bounds for the actual delay distribution in small cell networks.}.

\subsubsection{Random Scheduling} During each time slot, every active SAP randomly selects one of its UEs to serve.

\subsubsection{Round Robin} The SAPs arrange their UEs in a sequential order, and successively select one of the UEs to serve in each slot.

\newcounter{TempEqCnt}
\setcounter{equation}{\value{equation}}
\setcounter{equation}{5}
\begin{figure*}[t!]
\small
\begin{align} \label{equ: Meta_Grl}
F^{\Phi}(u) = \frac{1}{2} - \frac{1}{\pi} \int_0^\infty \frac{1}{\omega} \mathrm{Im}\left\{ u^{-j \omega} \left[ 1 + \delta \sum_{k=1}^{\infty} \binom{j \omega}{k} \left( F^\Phi( \xi K_{\mathrm{s}} ) + \int_{ \xi K_{\mathrm{s}} }^1 \frac{\left( \xi K_{\mathrm{s}} \right)^k}{t^k}  F^\Phi(dt) \right) \mathcal{Z}(k, \delta, \theta) \right]^{-1} \right\} d\omega
\end{align}
\normalsize
\setcounter{equation}{\value{equation}}{}
\setcounter{equation}{6}
\centering \rule[0pt]{18cm}{0.3pt}
\end{figure*}
\setcounter{equation}{\value{TempEqCnt}}
\setcounter{equation}{1}
\section{Analysis}

\subsection{Preliminaries}

\subsubsection{Service Rate}
Using the Slivnyak's theorem \cite{BacBla:09}, we can focus on a \textit{typical} {UE} who locates at the origin and receives data from its tagged SAP at $x_0$.
The corresponding service rate at time slot $t$ can then be written as
\setcounter{equation}{\value{equation}}
\setcounter{equation}{0}
\small
\begin{align}\label{equ: SIR}
\mu_{x_0, t}^\Phi = \mathbb{P}\left( \frac{P_{\mathrm{st}} h_{x_0} \Vert x_0 \Vert^{-\alpha} }{  \sum\limits_{x \in \Phi_{\mathrm{s}}\setminus x_0  } {P_{\mathrm{st}} \zeta_{x,t} h_{x}}{ \Vert x \Vert^{-\alpha}}   } > \theta \Bigg | \Phi_{\mathrm{s}} \right),
\end{align}
\normalsize
where $h_x$ is the channel fading from SAP $x$ to the origin, $\alpha$ is the path loss exponent, $\theta$ the SIR decoding threshold, and $\zeta_{x,t} \in \{0,1\}$ represents an indicator showing whether an SAP located at $x \in \Phi_{\mathrm{s}}$ is transmitting at time slot $t$ (in this case, $\zeta_{x,t} = 1$) or not (in this case, $\zeta_{x,t}=0$).
\setcounter{equation}{\value{equation}}
\setcounter{equation}{1}
\setcounter{equation}{\value{equation}}

Due to the broadcast nature of wireless medium, the queuing status of each SAP is coupled with other transmitters and hence results in interacting queue \cite{BerGalHum:92}. Consequently, the SAP active state, $\zeta_{x,t}$, is both spatial and temporal dependent: The spatial locations determine how the SAPs interfere with each other, and the temporal traffic dynamic affects the queue evolution.
Moreover, as a generic UE may see common interferencing SAPs in consecutive time slots, the system service rate is temporally correlated \cite{haenggi2016meta,GhaElsBad:17,chisci2017scalability}. Such temporal correlation introduces memory to the queue and can highly complicate the analysis.
In the following, we assume each link experiences independent service rate across time to maintain the mathematical tractability  \cite{GhaElsBad:17,chisci2017scalability}.

\subsubsection{Mean Delay} In this paper, we adopt the mean delay, i.e., the average number of required slots for one successful packet transmission, as our performance metric. A formal definition is given as follows.

\begin{definition}
\textit{
Let $A_x(T)$ be the number of packets arrived at a typical transmitter $x$ within period $[0,T]$, and $D_{i, x}$ be the number of time slots between the arrival of the $i$-th packet and its successful delivery. The mean delay is defined as
\small
\begin{align}\label{equ:Gen_Delay}
\mathbf{D}_x \triangleq    \lim\limits_{T \rightarrow \infty} \frac{\sum_{i=1}^{A_x(T)} D_{i, x} }{A_x(T)}.
\end{align}
\normalsize
}
\end{definition}
Note that $\mathbf{D}_x$ encapsulates the cross-network delay information as it is obtained by sampling at a random transmitter. Moreover, while $\mathbf{D}_x$ is defined via ergodic average, it is nevertheless a random variable because of the random service rate. Besides, the element $D_{i,x}$ in \eqref{equ:Gen_Delay} refers to the total amount of time that the $i$-th packet is in the queuing system, both in the queue (caused by other accumulated unsent packets) and in the service (due to link failure and retransmission).

\subsection{Delay Analysis}

This section details the main results of our work.
First of all, we condition on the random service rate, and give an explicit form of the mean delay for RS protocol.
\begin{lemma}\label{prop:Con_RS}
\textit{
Given the UE number $K_{\mathrm{s}}$, the arrival rate $\xi$, and the service rate $\mu^\Phi_{x_0}$, the mean delay at a typical SAP under RS is given by
\small
\begin{align}
\mathbf{D}_{x_0}^{\mathrm{RS}} = \left \{ \!\!\!
\begin{tabular}{cc}
$\frac{1-\xi}{ \frac{\mu^\Phi_{x_0}}{K_{\mathrm{s}}} - \xi}$, & if $\mu^\Phi_{x_0}/K_{\mathrm{s}} > \xi$,   \\
+$\infty$, &  if $\mu^\Phi_{x_0}/K_{\mathrm{s}} \leq \xi$.
\end{tabular}
\right.
\end{align}
\normalsize
and the queue non-empty probability is given by
\small
\begin{align}
\tau_a^{\mathrm{RS}} = \min\{ { K_{\mathrm{s}} \xi }/{\mu^\Phi_{x_0}}, 1\}.
\end{align}
\normalsize
}
\end{lemma}
\begin{IEEEproof}
See \cite{ZhoQueGe:16,yang2017packet} for a detail proof.
\end{IEEEproof}

The distribution of mean delay under RS protocol can be readily derived.
\begin{theorem}\label{thm:delay_RS}
\textit{
The cumulative distribution function (CDF) of the mean delay under RS is given by
\small
\begin{align}\label{equ: Delay_Dist_RS}
\mathbb{P}\left( \mathbf{D}_{x_0}^{\mathrm{RS}} \leq T \right) &= {1 - F^{\Phi} \left( \left( \frac{1-\xi}{T} + \xi \right) K_{\mathrm{s}} \right)}
\end{align}
\normalsize
where $F^{\Phi}(x)$ is given by the fixed-point equation \eqref{equ: Meta_Grl} on top of this page, with $\delta = 2 / \alpha$, $\mathrm{Im}\{\cdot\}$ being the imaginary part of a complex number, and $\mathcal{Z}(k, \delta, \theta)$ written as
\setcounter{equation}{\value{equation}}
\setcounter{equation}{6}
\small
\begin{align}
\mathcal{Z}(k, \delta, \theta) = \frac{ (-1)^{k+1} \theta^k }{k-\delta} {}_2F_1(k, k-\delta; k-\delta+1, -\theta),
\end{align}
\normalsize
whereas ${}_2F_1(a, b; c, d)$ is the hypergeometry function \cite{YanQue:18}.
\setcounter{equation}{\value{equation}}
\setcounter{equation}{7}
\setcounter{equation}{\value{equation}}
}
\end{theorem}
\begin{IEEEproof}
See \cite{YanQue:18} for a detail proof.
\end{IEEEproof}
The function given in \eqref{equ: Meta_Grl} can be solved via an iterative approach, and low-computational-complexity approximation is available to boost up the convergent speed \cite{YanQue:18}.

In regard to the RR protocol, we notice that each UE is scheduled to access the wireless channel in deterministic pattern. Hence, by conditioning on the service rate, we obtain an explicit form of mean delay as below.
\begin{lemma}\label{prop:Con_RR}
\textit{
Given the UE number $K_{\mathrm{s}}$, the arrival rate $\xi$, and the service rate $\mu^\Phi_{x_0}$, the mean delay at a typical SAP under RR is given as
\small
\begin{align}
\mathbf{D}_{x_0}^{\mathrm{RR}} = \left \{ \!\!\!\!\!\!\!\!\!\!
\begin{tabular}{cc}
& $\frac{1- {(K_{\mathrm{s}}+1) \xi}/{2}}{ {\mu^\Phi_{x_0}}/{K_{\mathrm{s}}} - \xi }$ - $\frac{K_{\mathrm{s}}-1}{2}$,  if $\mu^\Phi_{x_0}/K_{\mathrm{s}} > \xi$,   \\
&+$\infty$,  $\qquad \qquad \qquad$  if $\mu^\Phi_{x_0}/K_{\mathrm{s}} \leq \xi$.
\end{tabular}
\right.
\end{align}
\normalsize
and the queue non-empty probability is given by
\small
\begin{align}
\tau_a^{\mathrm{RR}} = \min\{ { K_{\mathrm{s}} \xi }/{\mu^\Phi_{x_0}}, 1\}.
\end{align}
\normalsize
}
\end{lemma}
\begin{IEEEproof}
See Appendix~\ref{apx:Con_RR} for a sketch of the proof.
\end{IEEEproof}
With the above result, the distribution of mean delay under RR can be derived subsequently.
\begin{theorem}\label{thm:delay_RR}
\textit{
The CDF of the mean delay under RR is given by
\small
\begin{align}\label{equ: Delay_Dist_RR}
\mathbb{P}\left( \mathbf{D}_{x_0}^{\mathrm{RR}} \leq T \right) &= {1 - F^{\Phi} \left( \left( \frac{ 1 - \frac{(K_{\mathrm{s}}+1)}{2} \xi }{T + \frac{K_{\mathrm{s}}-1}{2}} + \xi \right) K_{\mathrm{s}} \right)}
\end{align}
\normalsize
where $F^{\Phi}(x)$ is given by \eqref{equ: Meta_Grl}.
}
\end{theorem}
\begin{IEEEproof}
The proof is similar to that of Theorem~\ref{thm:delay_RS}.
\end{IEEEproof}
{\remark{\textit{ From Lemma~1 and Lemma~2, we note that when $\mu^\Phi_{x_0}/K_{\mathrm{s}} > \xi$,
there is $\mathbf{D}_{x_0}^{\mathrm{RS}} - \mathbf{D}_{x_0}^{\mathrm{RR}} \!=\! \frac{K_{\mathrm{s}} - 1 }{2} \frac{\mu^\Phi_{x_0}}{\mu^\Phi_{x_0} - K_{\mathrm{s}} \xi } > 0 $.
This observation confirms that RR outperforms RS in terms of delay regardless of traffic condition, which is in line with \cite{ZhoQueGe:16}.
 } }}
{\remark{\textit{ Take $T=1$ in the expressions of Theorem~1 and Theorem~2, we find that $\mathbb{P}\left( \mathbf{D}_{x_0}^{\mathrm{RS}} \leq 1 \right) = \mathbb{P}\left( \mathbf{D}_{x_0}^{\mathrm{RR}} \leq 1 \right) = 0$, i.e., the UEs that are able to success without retransmission form a probabilistic null set in small cell networks.
 } }}
{\remark{\textit{ When $T\gg1$, we have $\mathbb{P}\left( \mathbf{D}_{x_0}^{\mathrm{RS}} > T \right) \approx F^{\Phi}(\xi K_{\mathrm{s}}) + 1/T$ and $\mathbb{P}\left( \mathbf{D}_{x_0}^{\mathrm{RR}} > T \right) \approx F^{\Phi}(\xi K_{\mathrm{s}}) + 1/T$, which reveals that the CDF of mean delay obeys fat tail property.
 } }}

\section{Simulation and Numerical Results}
\begin{figure}[t!]
  \centering{}

    {\includegraphics[width=0.95\columnwidth]{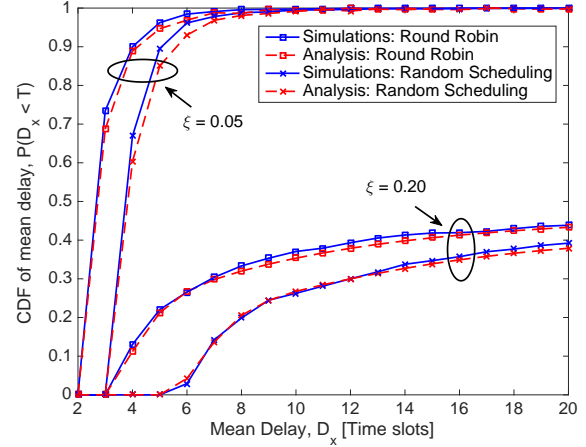}}

  \caption{CDF of mean delay: Simulations vs Analysis.}
  \label{fig:TDDPktThrPut}
\end{figure}

We now give simulation results that validate the accuracy of our analysis, and numerical results to compare the performance of the employed scheduling schemes.
Unless otherwise stated, we adopt the following system parameters \cite{yang2017packet,ZhoQueGe:16}: $\LS = 10^{-4} \mathrm{m}^{-2}$, $\PST = 23$~dBm, $K_{\mathrm{s}}=3$, $\theta = 0$~dB, $T_0 = 20$, and $\alpha = 3.8$.

In Fig.~\ref{fig:TDDPktThrPut}, we plot the CDF of mean delay under two different traffic statistics, namely, light traffic (where $\xi = 0.05$) and heavy traffic (where $\xi = 0.20$).
The figure shows a close match between simulations and analytical results, thus validates Theorem~1 and Theorem~2.
Moreover, it can be seen that RR attains smaller mean delay than RS in both light and heavy traffic regime, which is consistent with Remark~1.
Additionally, we note that the gain from RR is more pronounced under heavy traffic condition. Because when traffic profile grows, not only more newly arrived packets will build up the buffer, that incurs longer queuing delay, but more severely, more idle SAPs are activated and hence increases the interference, which results in longer transmission delay.
Under such circumstance, a scheduling protocol that regulates network inputs and grants each UE a fair service opportunity is critical for delay performance.
Hence, compared to RS that provides only random channel access, RR is able to achieve much better gain in delay via regulated access control.

\begin{figure}[t!]
  \centering{}

    {\includegraphics[width=0.95\columnwidth]{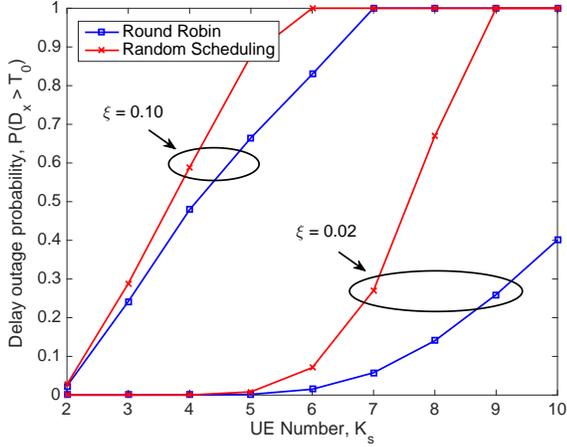}}

  \caption{Delay outage probability versus per-cell UE number.}
  \label{fig:Densification}
\end{figure}
Fig.~\ref{fig:Densification} depicts the delay outage probability, i.e., $\mathbb{P}(\mathbf{D}_{x} > T_0)$, as a function of UE number, $K_{\mathrm{s}}$, per cell.
This figure gives information about the scablility, which is crucial for internet-of-things (IoT) applications \cite{GhaElsBad:17}, of both scheduling protocols.
We can see that while RR possesses a better scability than RS, the relative performance varies with the traffic conditions, namely:
$i$) in the presence of light traffic, i.e., $\xi = 0.02$, RR can support much more UEs with relatively small delay outage probability than that of RS;
$ii$) in the medium to heavy traffic regime, i.e., $\xi = 0.10$, the delay outage probabilities under the two scheduling policies tends to be more alike as UE number grows.

\section{Conclusion}
In this paper, we conducted a study on the delay performance of small cell networks under random scheduling (RS) and round robin (RR) protocols.
For networks where both topology and traffic arrivals are random, we analyzed the mean delay by accounting for effects from queuing, retransmissions, and spatial-temporal interactions.
Our results confirmed that RR outperforms RS regardless of traffic condition, and the gain from RR is especially significant in the presence of heavy traffic, which demonstrates the importance of accounting fairness in the design of scheduling policy.
Moreover, we found that by using RR, the network can support larger amount of UEs with relatively small delay outage probability than that of RS, hence appeals RR as a suitable candidate for scheduling protocols in IoT applications.

\begin{appendix}

\subsection{Sketch of Proof of Lemma~\ref{prop:Con_RR}} \label{apx:Con_RR}
Under round robin, a typical UE is scheduled for transmission in every $K_{\mathrm{s}}$ slots.
During the time slots when the UE is not scheduled, there is only packet arrivals, and the queue state transitioin matrix is given by
\small
\begin{align}
\mathbf{P}_{\mathrm{A}} \!=\!
  \begin{bmatrix}
    ~1 - \xi & \!\!\!\!\! \xi &\!\!  0 &\!\!\!  \cdots &\!\!\!\!\! \cdots &\!\!\!\!\! \cdots &\!\!\! 0 &\!\!\! 0 &\!\!\! \cdots \
    \nonumber\\
    ~0 & \!\!\!\!\! 1 - \xi &\!\!\!  \xi &\!\!\!  0 &\!\!\!\!\! \cdots &\!\!\!\!\! \cdots &\!\!\! 0 &\!\!\! 0 &\!\!\! \cdots \
    \nonumber\\
    ~\vdots & \!\!\!\!\!\vdots &\!\!\!  \vdots &\!\!\! \vdots &\!\!\!\!\! \ddots &\!\!\!\!\!  \cdots & \!\!\! 0 &\!\!\! 0 &\!\!\! \cdots \
    \nonumber\\
    ~0 & \!\!\!\!\!0 &\!\!\!  0 &\!\!\! 0 &\!\!\!\!\! \cdots &\!\!\!\!\! 1 - \xi &\!\!\! \xi &\!\!\! 0 &\!\!\! \cdots\
    \nonumber\\
    ~\vdots & \!\!\!\!\! \vdots &\!\!\! \vdots &\!\!\! \vdots &\!\!\!\!\! \vdots &\!\!\!\!\! \vdots &\!\!\! \vdots &\!\!\! \vdots &\!\!\! \ddots \
   \end{bmatrix}.
\end{align}
\normalsize
During the scheduled transmission slots, the queue state transition matrix can be written as follows
\small
\begin{align}
\mathbf{P}_{\mathrm{D}} \!=\!
  \begin{bmatrix}
    ~ 1-\xi & \!\!\!\!\! \xi &\!\!\!\!\! 0 &\!\!\!\!\! \cdots &\!\!\!\!\! \cdots &\!\!\!\!\! \cdots &\!\!\!\!\! 0 &\!\!\!\!\!0 &\!\!\!\!\!\cdots \
    \nonumber\\
    ~\kappa_2 & \!\!\!\!\! \kappa_3 &\!\!\!\!\! \kappa_1 &\!\!\!\!\! 0 &\!\!\!\!\! \cdots &\!\!\!\!\! \cdots &\!\!\!\!\! 0 &\!\!\!\!\!0 &\!\!\!\!\!\cdots \
    \nonumber\\
    ~0 &\!\!\!\!\! \kappa_2 & \!\!\!\!\! \kappa_3 &\!\!\!\!\! \kappa_1 &\!\!\!\!\! 0 &\!\!\!\!\! \cdots &\!\!\!\!\!  0 &\!\!\!\!\! 0 &\!\!\!\!\!\cdots \
    \nonumber\\
    ~\vdots & \!\!\!\!\!\vdots &\!\!\!\!\! \vdots &\!\!\!\!\! \vdots &\!\!\!\!\! \ddots &\!\!\!\!\!  \cdots & \!\!\!\!\! 0 &\!\!\!\!\! 0 &\!\!\!\!\! \cdots \
    \nonumber\\
    ~0 & \!\!\!\!\!0 &\!\!\!\!\! 0 &\!\!\!\!\! 0 &\!\!\!\!\! \cdots &\!\!\!\!\! \kappa_2 &\!\!\!\!\! \kappa_3 &\!\!\!\!\! \kappa_1 &\!\!\!\!\! \cdots\
    \nonumber\\
    ~\vdots & \!\!\!\!\! \vdots &\!\!\!\!\! \vdots &\!\!\!\!\! \vdots &\!\!\!\!\! \vdots &\!\!\!\!\! \vdots &\!\!\!\!\! \vdots &\!\!\!\!\! \vdots &\!\!\!\!\! \ddots \
   \end{bmatrix}
\end{align}
\normalsize
where $\kappa_1$, $\kappa_2$, and $\kappa_3$ are respectively given by
\small
\begin{align}
\kappa_1 &= \xi ( 1- \mu^\Phi_{x_0}), \quad \kappa_2 = (1 - \xi ) \mu^\Phi_{x_0},\\
\kappa_3 &= (1 - \xi) (1 - \mu^\Phi_{x_0}) + \xi \mu^\Phi_{x_0}.
\end{align}
\normalsize

Without loss of generality, we let the timing start from the moment the UE just been scheduled, hence, the queue transition matrix in a complete transmission round can be written as
\small
\begin{align}
\mathbf{P}_{\mathrm{T}} =
\underbrace{ \mathbf{P}_{\mathrm{A}} \times \mathbf{P}_{\mathrm{A}} \times \cdots \times \mathbf{P}_{\mathrm{A}} }_{K_{\mathrm{s}} -1 }  \times \mathbf{P}_{\mathrm{D}}
\end{align}
\normalsize
where the product is performed with respect to the regular matrix multiplication.
Let $\mathbf{v} = \left( v_1, v_2, ... \right)$ denote the steady state probability vector, then, in the steady state, we have
\small
\begin{align}{\label{equ:treq}}
\mathbf{v} \mathbf{P}_{\mathrm{T}} &= \mathbf{v}, \\
\sum_{i=0}^\infty v_i &= 1.
\end{align}
\normalsize
Solving the above system equation yields the individual elements $\{v_i\}_{i=0}^\infty$, the queue non-empty probability corresponds to $\tau^{\mathrm{RR}}_{x_0} = 1-v_0$, and the mean delay can then be obtained as
\small
\begin{align}
\mathbf{D}_{x_0}^{\mathrm{RR}} = \sum_{i=0}^\infty i v_i + \frac{1}{\mu^\Phi_{x_0}}.
\end{align}
\normalsize
\end{appendix}

\bibliographystyle{IEEEtran}
\bibliography{bib/StringDefinitions,bib/IEEEabrv,bib/howard_trff_SCN}

\end{document}

%% file: SupportDocuments/aconym.tex
\acrodef{CCDF}{complementary cumulative distribution function}
\acrodef{CF}{characteristic function}
\acrodef{PPP}{Poisson point processe}
\acrodef{RV}{random variable}
\acrodef{i.i.d.}{independent and identically distributed}
\acrodef{PDF}{probability distribution function}
\acrodef{CDF}{cumulative distribution function}
\acrodef{ch.f.}{characteristic function}
\acrodef{AWGN}{additive white Gaussian noise}
\acrodef{SNR}{signal-to-noise ratio}
\acrodef{LRT}{likelihood ratio test}
\acrodef{DRT}{distance ratio test}
\acrodef{GLRT}{generalized likelihood ratio test}
\acrodef{CRLB}{Cram\'{e}r-Rao lower bound}
\acrodef{CRB}{Cram\'{e}r-Rao bound}
\acrodef{ZZLB}{Ziv-Zakai lower bound}
\acrodef{ZZB}{Ziv-Zakai bound}
\acrodef{LOS}{line-of-sight}
\acrodef{ToF}{time-of-flight}
\acrodef{NLOS}{non-line-of-sight}
\acrodef{GDOP}{geometric dilution of precision}
\acrodef{GPS}{Global Positioning System}
\acrodef{FIM}{Fisher information matrix}
\acrodef{PEB}{position error bound}
\acrodef{SPEB}{squared position error bound}
\acrodef{TOA}{time-of-arrival}
\acrodef{TOF}{time-of-flight}
\acrodef{WSN}{wireless sensor network}
\acrodef{MAC}{medium access control}
\acrodef{RSS}{received signal strength}
\acrodef{WAF}{wall attenuation factor}
\acrodef{TDOA}{time difference-of-arrival}
\acrodef{RF}{radiofrequency}
\acrodef{RTT}{round-trip time}
\acrodef{AOA}{angle-of-arrival}
\acrodef{MF}{matched filter}
\acrodef{ED}{energy detector}
\acrodef{ML}{maximum likelihood}
\acrodef{MSE}{mean-square error}
\acrodef{RMSE}{root-mean-square error}
\acrodef{LEO}{localization error outage}
\acrodef{ppm}{part-per-million}
\acrodef{ACK}{acknowledge}
\acrodef{UWB}{Ultrawide bandwidth}
\acrodef{TNR}{threshold-to-noise ratio}
\acrodef{LS}{least squares}
\acrodef{IR-UWB}{impulse radio UWB}
\acrodef{FCC}{Federal Communications Commission}
\acrodef{TH}{time-hopping}
\acrodef{PPM}{pulse position modulation}
\acrodef{MUI}{multi-user interference}
\acrodef{PDP}{power delay profile}
\acrodef{BPZF}{band-pass zonal filter}
\acrodef{SIR}{signal-to-interference ratio}
\acrodef{SINR}{signal-to-interference-plus-noise ratio}
\acrodef{RFID}{radio frequency identification}
\acrodef{WPAN}{wireless personal area network}
\acrodef{WWB}{Weiss-Weinstein bound}
\acrodef{DP}{direct path}
\acrodef{MF}{matched filter}
\acrodef{MMSE}{minimum-mean-square-error}
\acrodef{SBS}{serial backward search}
\acrodef{SBSMC}{serial backward search for multiple clusters}
\acrodef{NBI}{narrowband interference}
\acrodef{WBI}{wideband interference}
\acrodef{INR}{interference-to-noise ratio}
\acrodef{CR}{channel response}
\acrodef{CIR}{channel impulse response}
\acrodef{CR}{channel  response}
\acrodef{RADAR}{radar}
\acrodef{MUR}{Multistatic radar}
\acrodef{JBSF}{jump back and search forward}
\acrodef{HDSA}{high-definition situation-aware}
\acrodef{RRC}{root raised cosine}
\acrodef{ST}{simple thresholding}
\acrodef{BTB}{Bellini-Tartara bound}
\acrodef{P-Max}{$P$-Max}  
\acrodef{MIMO}{multiple-input multiple-output}
\acrodef{MAP}{maximum a posteriori}
\acrodef{FG}{factor graph}
\acrodef{OP}{outage probability}
\acrodef{WED}{wall extra delay}
\acrodef{RMS}{root mean square}
\acrodef{SPAWN}{sum-product algorithm over a wireless network}
\acrodef{MDD}{minimum distance distribution}
\acrodef{MAP}{maximum a posteriori probability}
\acrodef{SAP}{small cell access point}
\acrodef{UE}{user equipment}
\acrodef{MBS}{macro cell base station}
\acrodef{UER}{\ac{UE} Relay}
\acrodef{D2D}{device-to-device}
\acrodef{MBS}{macro base station}
\acrodef{CSI}{channel state information}
\acrodef{OGR}{outage guard region}
\acrodef{FUR}{feasible UER region}
\acrodef{EHR}{energy harvesting region}
\acrodef{EH}{energy harvesting}
\acrodef{D2D-EHSN}{D2D communication provided \ac{EH} small cell network}
\acrodef{D2D-EHHN}{D2D communication provided \ac{EH} heterogeneous network}
\acrodef{3GPP}{3rd Generation Partnership Project}
\acrodef{BS}{base station}
\acrodef{DF}{decode and forward}
\acrodef{CCDF}{complementary cumulative distribution function}
\acrodef{ZF}{zero forcing}
\acrodef{RZF}{regularized zero forcing}
\acrodef{WLLN}{weak law of large number}
\acrodef{SLLN}{strong law of large numbers}
\acrodef{TDD}{Time-division duplex}
\acrodef{EE}{energy efficiency} 
\acrodef{HetNet}{heterogeneous network} 
\acrodef{SCP}{Single Cell Processing}
\acrodef{CBF}{Coordinated Beamforming}

%% file: SupportDocuments/defmetric.tex
\usepackage{color}
\usepackage{dsfont}
\usepackage{bbm}

\def\PST{P_{\mathrm{st}}}

\def\LS{\lambda_{\mathrm{s}}}

\DeclareMathAlphabet{\mathsf}{OML}{cmbr}{m}{it}

\newtheorem{definition}{\bf Definition}
\newtheorem{theorem}{\bf Theorem}
\newtheorem{lemma}{\bf Lemma}

\newtheorem{remark}{\bf Remark}

\newcommand{\bd}{\begin{description}}
\newcommand{\ed}{\end{description}}
\newcommand{\be}{\begin{enumerate}}
\newcommand{\ee}{\end{enumerate}}
\newcommand{\bi}{\begin{itemize}}
\newcommand{\ei}{\end{itemize}}
\newcommand{\bl}{\begin{list}}
\newcommand{\el}{\end{list}}
\newcommand{\bt}{\begin{tabbing}}
\newcommand{\et}{\end{tabbing}}